\documentclass[letterpaper,twocolumn,showpacs,pra,aps]{revtex4-1} 
\usepackage{amsmath}  
\usepackage{amsfonts} 
\usepackage{graphicx} 
\usepackage{amssymb}

\usepackage{marginnote}
\usepackage{xcolor}

\begin{document}

\title{Kelvin-Froude wake patterns of a traveling pressure disturbance}

\author{Jonathan Colen$^{1}$ and Eugene B. Kolomeisky$^{2}$}

\affiliation
{$^{1}$Department of Physics, University of Chicago, 5720 South Ellis Ave, Chicago, Illinois 60637, USA\\
$^{2}$Department of Physics, University of Virginia, P. O. Box 400714,
Charlottesville, Virginia 22904-4714, USA}

\date{\today}

\begin{abstract}
According to Kelvin, a point pressure source uniformly traveling over the surface of deep calm water leaves behind universal wake pattern confined within $39^{\circ}$ sector and consisting of the so-called transverse and diverging wavefronts.  Actual ship wakes differ in their appearance from both each other and Kelvin's prediction.  The difference can be attributed to a deviation from the point source limit and for given shape of the disturbance quantified by the Froude number $F$.  We show that within linear theory effect of arbitrary disturbance on the wake pattern can be mimicked by an effective pressure distribution.  Further, resulting wake patterns are qualitatively different depending on whether water-piercing is present or not ("sharp" vs "smooth" disturbances).   For smooth pressure sources, we generalize Kelvin's stationary phase argument to encompass finite size effects and classify resulting wake patterns. Specifically, we show that there exist two characteristic Froude numbers, $F_{1}$ and  $F_{2}>F_{1}$, such as the wake is only present if $F\gtrsim F_{1}$.  For $F_{1}\lesssim F \lesssim F_{2}$, the wake consists of the transverse wavefronts confined within a sector of an angle that may be smaller than Kelvin's.  An additional $39^{\circ}$ wake made of both the transverse and diverging wavefronts is found for $F\gtrsim F_{2}$.  If the pressure source has sharp boundary, the wake is always present and features additional interference effects.  Specifically, for a constant pressure line segment source mimicking slender ship the wake pattern can be understood as due to two opposing effect wakes resembling (but not identical to) Kelvin's and originating at segment's ends.  

\end{abstract}


\maketitle

\section{Motivation}

\begin{figure*}[!t]
\centering
\includegraphics[width=2.0\columnwidth, keepaspectratio]{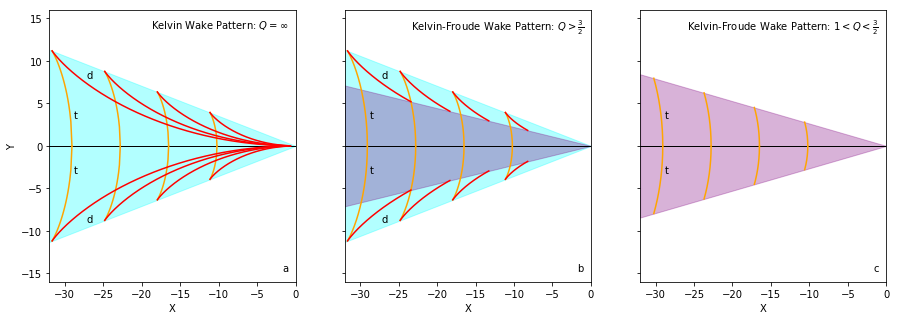} 
\caption{(Color online) Evolution of the wavecrests produced by pressure source at the origin traveling to the right, Eqs.(\ref{parametric}), (\ref{wavecrests}) and (\ref{range}), in co-moving reference frame as a function of the cutoff parameter $Q$: from Kelvin wake (a) to Kelvin-Froude wakes, (b) and (c).  Kelvin units of length (\ref{scale}) are adopted hereafter.  Transverse (t) and diverging (d) wavefronts are shown in orange and red, respectively.  Regions where both transverse and diverging wavefronts are found are shaded light blue;  regions with only transverse wavefronts present are shaded grey (b) or light purple (c).  In cases (a) and (b) the entire wake is bounded by Kelvin's $39^{\circ}$ angle.  Opening angle of the inner sector with only transverse wavefronts present, cases (b) and (c), is given by Eq.(\ref{inner_angle}).}
\label{wakes}
\end{figure*}

It is impossible to overlook similarity of the wakes produced on deep water by objects as distinct in sizes, shapes and speeds as a waterfowl, a high-speed boat, or a tanker.  This property has its origin in hydrodynamic similarity of the flow pattern due to a point traveling pressure source as discovered by Kelvin \cite{Kelvin}.  The conclusion is a consequence of linearity of the theory and the dispersion law of gravity waves on deep water, which in the inviscid incompressible fluid limit, and neglecting the effects of capillarity, has the form \cite{Lamb}
\begin{equation}
\label{spectrum}
\omega^{2}(\textbf{k})=gk
\end{equation}
where $\omega$ is the frequency of the wave of the wave vector $\textbf{k}$, $g$ is the free fall acceleration and $k=|\textbf{k}|$.  Indeed, dimensional considerations imply that the flow pattern is characterized by a single length scale
\begin{equation}
\label{scale}
l=\frac{v^{2}}{g},
\end{equation}
hereafter called the Kelvin length. Rescaling lengths by $l$ yields a parameter-free problem, demonstrating the geometric similarity of wakes produced by point sources. The resulting Kelvin wake (Figure \ref{wakes}a) has a striking "feathered" appearance caused by transverse and diverging wavefronts confined within a $39^{\circ}$ sector~\cite{Kelvin, Lamb, Newman, Faber}. 

The Kelvin wake is a dispersive counterpart of the Cherenkov and Mach wakes \cite{wake_review} produced by light and sound, respectively.
However, its formation does not require the source exceeding a minimum speed. Indeed, the wake is present if there is a wave mode whose phase velocity $\omega/k$ matches the projection of the velocity of the source onto the direction of radiation.  This statement, known in hydrodynamics as the condition of stationarity \cite{Newman,Faber} or elsewhere in physics as the Mach-Cherenkov-Landau (MCL) constraint \cite{wake_review}, applied to the dispersion law (\ref{spectrum}) acquires the form
\begin{equation}
\label{Cherenkov}
\cos\varphi=\frac{\omega}{kv}=\sqrt{\frac{g}{v^{2}k}}\equiv\frac{1}{\sqrt{kl}}
\end{equation}     
where $\varphi$ is the angle between the direction of radiation of the gravity wave and the direction of motion of the source.  Since $\cos\varphi \leqslant 1$, the wake is present if the inequality $kl\geqslant1$ holds \cite{NHY}.  Clearly there always are modes satisfying this condition and thus participating in producing the wake.     

Ship wakes seen in practice are not strictly similar.  Indeed, the wakes of tankers are dominated by transverse wavefronts while the wakes of high-speed boats are largely made of diverging waves. This is due to finite-size effects neglected by Kelvin~\cite{Lighthill}.  These effects are characterized by the Froude number $F$ relating (for fixed shape) the characteristic length $a$ of the source (e.g. the hull length) to the Kelvin length (\ref{scale})
\begin{equation}
\label{Froude}
F=\sqrt{\frac{l}{a}}=\frac{v}{\sqrt{ga}},
\end{equation}
The Froude number, which also is central to understanding wave resistance~\cite{Lamb,Newman,Faber}, is a relative of the Mach number
\begin{equation}
\label{Mach}
M=\frac{v}{c}
\end{equation}   
where the speed of sound (or light) $c$ is the counterpart of the combination $\sqrt{ga}$ also known as the hull speed \cite{Falkovich}.  

In marine practice, the applicability of linear water wave theory~\cite{Lamb} is limited by wave breaking, which sets an upper bound on the Froude number $F \simeq 3$ that can realistically be achieved by high-speed boats. Thus, deviations from the ideal Kelvin wake ($a=0$ or $F=\infty$) are not surprising.

The apparent wake angle formed by directions of the peaks of highest waves is a direct signature of the wake pattern, and ship wakes narrower than Kelvin's have long been observed both in practice \cite{observations1,observations2,observations3,observations4,observations5,observations6} and numerical studies \cite{numerical1,numerical2,numerical3}.  
The interest in ship wake appearances has been reignited following a study of Rabaud and Moisy~\cite{RM}, who used Google Earth images of ship wakes to examine how the wake angle depends on the Froude number. They found the angle approximately constant and close to Kelvin's $39^{\circ}$ prediction for $0.1\lesssim F \lesssim 0.6$. As $F$ increased beyond this range, the angle decreased, reaching values as small as $14^{\circ}$ at $F\simeq 1.7$ seemingly calling Kelvin's theory into question.  These authors also argued that for $F$ large the wake angle obeys Mach's law, scaling as $1/F$. Their explanation of these observations assumed that an object cannot generate waves with wavelength larger than its size~\cite{RM}.

Darmon, Benzaquen, and Rapha\"el \cite{DBR} and Ellingsen \cite{Ellingsen} (who additionally accounted for the effects of shear currents of constant vorticity) provided an analysis of the observations \cite{RM} based on the classical linear water wave theory \cite{Lamb,theory,RD} applied to the case of traveling isotropic Gaussian pressure source.  While explicitly demonstrating that the angle of largest waves follows Mach's $1/F$ scaling for $F$ large, these treatments also indicated that the entire wake was delimited by Kelvin's $39^{\circ}$ angle.   Later analysis~\cite{BDR} extended to wake patterns due to anisotropic pressure disturbances. Moisy and Rabaud conducted a similar investigation~\cite{MR}, although their interpretation assumed that for a disturbance of size $a$, the excited wave amplitudes are large only for wavelengths in a narrow band around $a$.

Noblesse and collaborators~\cite{interference,YZhu} provide a different explanation of narrow wake angles which is consistent with reported observations. 
Their analysis extends a classic idea of marine hydrodynamics that a ship wake pattern can be seen as the interference of two opposite-effect Kelvin wakes originating from the bow and stern~\cite{Newman,Faber,resistance}. This approach predicts that for a slender ship, the angle of the largest waves scales as $1/F^2$ for large $F$. By representing the flow around ships as a continuous distribution of sources and sinks over the hull surface, the authors were able to demonstrate relevance of their ideas to realistic settings such as monohull ships and catamarans~\cite{Zhang,JHe,FNoblesse,anotherZhu,Wu}.

While assessing the assumptions underlying these three competing explanations of narrow wakes, He \textit{et al.}~\cite{He} cast skepticism on the conjectures constraining wavelengths generated by a ship~\cite{RM}. Although analysis of the wake pattern due to an isotropic Gaussian pressure distribution \cite{DBR,Ellingsen} seemed to justify these conjectures, the water flow due to this distribution does not properly represent that of a ship. Furthermore,  a generalization to anisotropic pressure sources \cite{BDR} still does not account for rapid variation of the flow near a ship's bow and stern.       

Submerged sources traveling parallel to the water surface also exhibit wake patterns deviating from Kelvin's prediction. For a point source, the deviation is characterized by the Froude number (\ref{Froude}) where $a$ is now the submergence depth. Analysis of this problem~\cite{Peti,AK,Wu} and that of a submerged force doublet~\cite{Peti} yield similar results as those for an isotropic Gaussian pressure source~\cite{DBR} at the free water surface. The common theme is that the wake pattern is dominated by the diverging waves for $F$ large and by the transverse ones for $F$ small.  Wu \textit{et al.} \cite{Wu} went even further and investigated a combined effect of the submergence depth and the hull length for a fully submerged slender body.           

These studies demonstrate the dramatic influence finite-size effects can have on ship waves, while stimulating fundamental questions regarding gravity wakes.  In this work, we will address these questions in a unified manner free of unnecessary assumptions. 

(i)  What is the physics behind the wakes more narrow than Kelvin's?  
Conjectures relating the excited wavelength and source size~\cite{RM,MR} are controversial~\cite{He} despite some analysis seeming to support them~\cite{DBR,BDR}. On the other hand, while observational evidence~\cite{Newman} supports the claim that the effect is primarily due to two-point interference~\cite{interference,YZhu}, this has not been proven theoretically.  The issue is that interfering diverging and transverse wavefronts generated by the bow and stern of the ship are not plane waves.

(ii)  Since there is more to the wake than apparent wake angle, how does the Froude number $F$ control the appearance of the wake? What are the possible wake patterns? There exist qualitative arguments explaining the role of the Froude number~\cite{Lighthill}, but no first-principles theoretical understanding has been developed.

(iii)  How is it possible for the wake to be bounded by Kelvin's $39^{\circ}$ angle for small $F$? It is a puzzle because in this case the pattern has to be least similar to the ideal, $F=\infty$, Kelvin wake.  

The answers to these questions lie in the observation that  
water piercing has a dramatic effect on wake patterns.  The same fact is also partly 
responsible for differences in predictions due to Rapha\"el \textit{et al.} \cite{DBR,BDR} and Noblesse \textit{et al.} \cite{interference}.

In this study we limit ourselves to linear potential flow theory of ideal liquid which is known to be accurate in most practical applications that involve ship waves \cite{Lamb,Newman,Faber,Lighthill,Falkovich}.  We also assume that deep water limit of the spectrum of gravity waves (\ref{spectrum}) holds.      

\section{Statement of the problem}

Like in Refs.\cite{RD,DBR,BDR}, we begin with an expression for the vertical displacement of the water surface due to a pressure source traveling with constant velocity $\textbf{v}$ over deep water in the co-moving reference frame \cite{Lamb,theory,RD}, 
\begin{equation}
\label{general_wake_integral}
\zeta(\textbf{r})=\frac{1}{\rho}\int\frac{d^{2}k}{(2\pi)^{2}}\frac{kp(\textbf{k})e^{i\textbf{k}\cdot\textbf{r}}}{(\textbf{k}\cdot\textbf{v}+i0)^{2}-gk},
\end{equation} 
where $\textbf{r}$ is a two-dimensional position vector, $\rho$ is the density of water, and $p(\textbf{k})$ is a Fourier transform of the excess pressure $\delta p(\textbf{r})$ due to traveling disturbance,
\begin{equation}
\label{Fourier_transform_definition}
p(\textbf{k})=\int \delta p(\textbf{r})e^{-i\textbf{k}\cdot \textbf{r}}d^{2}r.
\end{equation}

\subsection{Linear theory of ship waves is a linear response theory}

At first it appears that Eq.(\ref{general_wake_integral}) limits us to disturbances in the form of pressure distributions applied at free water surface and excludes water-piercing or underwater perturbations from consideration.  We however contend that representation of the wake in the form of the singular Fourier integral (\ref{general_wake_integral}) is completely general if $p(\textbf{k})$ is interpreted more broadly as a source or an \textit{effective} pressure function that can also mimic effects of a hull of a ship or an underwater object.

Linearity of the theory dictates that the water surface displacement 
is a superposition of plane waves $\exp(i\textbf{k}\cdot\textbf{r})$ with amplitudes proportional to the perturbation $p(\textbf{k})$, 
which partly explains the structure of Eq.(\ref{general_wake_integral}).  To understand the remainder of the integrand, note that Eq.(\ref{general_wake_integral}) originates from the equation of motion for the displacement $\zeta(\mathbf{k}, t)$~\cite{AK}: 
\begin{equation}
\label{2nd_law}
\frac{\rho}{k}\frac{d^{2}\zeta(\textbf{k},t)}{dt^{2}}=-\rho g \zeta(\textbf{k},t)-p(\textbf{k},t)
\end{equation}
The $\rho/k$ coefficient in the left-hand side gives the mass of water (per unit area) in motion, as the surface displacement $\zeta(\mathbf{k}, t)$ affects surface layer of depth of order $1/k$. The right-hand side includes both a restoring and external force (both per unit area). Thus, Eq.(\ref{2nd_law}) describes a forced harmonic oscillator of natural frequency $\omega(\mathbf{k})$ (\ref{spectrum}). 

Any time-dependent perturbation can be Fourier-expanded to a set of monochromatic components with time dependence $e^{-i \omega t}$.  Substituting in (\ref{2nd_law}) $p$ and $\zeta$ in the forms $p_{\omega}e^{-i\omega t}$ and $\zeta_{\omega}e^{-i\omega t}$ we find a relationship between the Fourier components of the force and the response
\begin{eqnarray}
\label{linear_response}
\zeta_{\omega}(\textbf{k})&=&-\alpha(\omega,\textbf{k})p_{\omega}(\textbf{k})\nonumber\\
\alpha(\omega,\textbf{k})&=&\frac{k}{\rho[\omega^{2}(\textbf{k})-(\omega+i0)^{2}]}
\end{eqnarray}
where we used Eq.(\ref{spectrum}).  In the language of the linear response theory the function $\alpha(\omega,\textbf{k})$ is the susceptibility of the system, and a shift $\omega\rightarrow\omega+i0$ enforces causality \cite{LL5}.

The Fourier component $p_{\omega}(\textbf{k})$ of a traveling disturbance, $\delta p(\textbf{r},t)=\delta p (\textbf{r}-\textbf{v}t)$, is related to its spatial counterpart (\ref{Fourier_transform_definition}) as $p_{\omega}(\textbf{k})=2\pi \delta(\omega-\textbf{k}\cdot\textbf{v})p(\textbf{k})$.
Inverting Eq.(\ref{linear_response}) to real space and time and changing 
to the source reference frame, $\textbf{r}-\textbf{v}t\rightarrow \textbf{r}$, we obtain
\begin{equation}
\label{real_linear response}
\zeta(\textbf{r}) =-\int\frac{d^{2}k}{(2\pi)^{2}}\alpha(\textbf{k}\cdot\textbf{v},\textbf{k})p(\textbf{k})e^{i\textbf{k}\cdot\textbf{r}}.
\end{equation} 
While this is identical to Eq.(\ref{general_wake_integral}), representation that uses the language of the linear response theory with the system's susceptibility $\alpha$ modulating the contribution of each plane wave $\exp(i\textbf{k}\cdot\textbf{r})$ into the water displacement makes it clear that Eq.(\ref{general_wake_integral}) is completely general.

By interpreting $p(\mathbf{k})$ as an effective pressure distribution which can mimic effects of a ship hull or underwater objects, we can also use this general Fourier integral to consider effects of water-piercing or underwater perturbations. 

Determination of the effective pressure function requires separate analysis tailored to each situation, as it depends on parameters such as the source type, shape, size, and speed.  For example, for the source function
\begin{equation}
\label{ball}
p(\textbf{k})=2\pi \rho R^{3} \frac{(\textbf{k}\cdot\textbf{v})^{2}}{k}e^{-ka}, 
\end{equation}       
the integral (\ref{general_wake_integral}) gives water surface elevation due to submerged ball of radius $R$ that travels with constant velocity $\textbf{v}$ at a depth $a\gg R$ \cite{theory,ball,AK}.     

\subsection{Connection to Fourier-Kochin representation of the wake}

We now compare this approach with the classic Fourier-Kochin representation of the \textit{far-field} part of the wake \cite{FK1,FK2,FK3,FK4,NHY,CZhang,JHe,anotherZhu,HWu1,HWu2,Wu}
\begin{equation}
\label{FK}
\zeta_{w}(\textbf{r})=\frac{1}{\pi}\Re \int_{-\infty}^{\infty}dqA(q) e^{i\sqrt{1+q^{2}}(x+qy)}
\end{equation}
Hereafter, length is rescaled by the Kelvin length (\ref{scale}) and $q^2 = k-1$. The Kochin function $A(q)$ contains information about the wake source using the convention of Ref.~\cite{Wu}.

\begin{figure}
\begin{center}
\includegraphics[width=1\columnwidth,keepaspectratio]{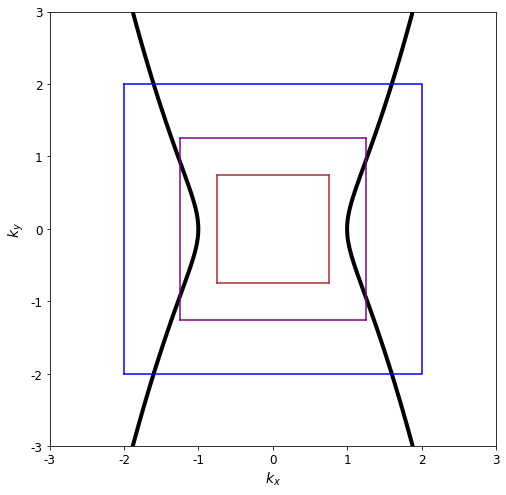}
\caption{(Color online)  Locus of the wave vectors whose components satisfy the MCL condition (\ref{pole}).  Colored overlaid squares represent boundaries of the $2k_{max}\times2k_{max}$ integration regions in (\ref{Kelvin_integral}) for three representative choices of $k_{max}$.}
\label{MCL}
\end{center}
\end{figure}

Defining the $+x$ direction along the direction of source motion in Eq.~(\ref{general_wake_integral}) yields
\begin{eqnarray}
\label{Kelvin_integral}
\zeta(\textbf{r})&=&\int\frac{d^{2}k}{(2\pi)^{2}}\frac{kp(\textbf{k})e^{i\textbf{k}\cdot\textbf{r}}}{(k_{x}+i0)^{2}-k}
=\int\frac{\sqrt{k}d^{2}k}{2(2\pi)^{2}}p(\textbf{k})e^{i\textbf{k}\cdot\textbf{r}}\nonumber\\
&\times&\left (\frac{1}{k_{x}-\sqrt{k}+i0}-\frac{1}{k_{x}+\sqrt{k}+i0}\right )
\end{eqnarray}
If the function $p(\textbf{k})$ is not too localized, the integral 
is dominated by the poles of the integrand
\begin{equation}
\label{pole}
k_{x}^{2}=k
\end{equation}
which is the MCL condition (\ref{Cherenkov}) in disguise. A locus of these points is shown in Figure \ref{MCL}. 
From the viewpoint of the equation of motion (\ref{2nd_law}), the MCL constraint (\ref{pole}) is the condition of resonant excitation of gravity waves by the external source.

The contribution of the poles to the integral corresponds to the far-field part of the wake and can be separated out using the formula \cite{LL5}
\begin{equation}
\label{principal}
\frac{1}{z-b+i0}=P\frac{1}{z-b}-i\pi \delta(z-b)
\end{equation}
Here, $P$ denotes the taking of the principal value when integrating the function $\phi(z)/(z-b)$.  Using Eq.(\ref{principal}) to integrate over $k_{x}$, the far-field part of the wake (\ref{Kelvin_integral}) can be written as
\begin{eqnarray}
\label{far-field}
\zeta_{w}(\textbf{r})&=&\frac{1}{2\pi} \Im \int_{-\infty}^{\infty}dq (1+q^{2})p\left (\sqrt{1+q^{2}},q\sqrt{1+q^{2}}\right )\nonumber\\
&\times& e^{i\sqrt{1+q^{2}}(x+qy)}
\end{eqnarray}
where $\Im$ stands for the imaginary part.  In arriving at this expression we employed the property that $p^{*}(\textbf{k})=p(-\textbf{k})$ (since $\delta p(\textbf{r})$ is real)
and assumed that the source is reflection-symmetric, $p(k_{x},k_{y})=p(k_{x},-k_{y})$.
Furthermore, we dropped the near-field contribution given by the principal value of the integral (\ref{general_wake_integral}) which rapidly falls off as one moves away from the source. The representations (\ref{far-field}) and (\ref{FK}) are essentially equivalent. 

\section{Kelvin wake pattern}

Representation of the wake pattern using the language of the linear response theory (\ref{real_linear response}) implies that for a point source $p(\textbf{k})=const$ the susceptibility $\alpha(\textbf{k}\cdot\textbf{v},\textbf{k})$ is essentially a Fourier transform of the original Kelvin wake.  Thus, the Kelvin limit $p(\textbf{k})=const$ continues to play central role for general $p(\textbf{k})$, and we review this case first.  

While the Fourier integral (\ref{Kelvin_integral}) cannot be computed in closed form even if $p(\textbf{k})=const$, the geometry of the wake pattern can be understood by employing Kelvin's stationary phase argument \cite{Kelvin,Lamb,Newman,Faber,wake_review}.

Far from the source $r\gg1$, the phase factor $f=\mathbf{k}\cdot\mathbf{r}$ is large and the exponential is highly oscillatory. Contributions from this region typically interfere destructively, leaving almost no net result. However, wave vectors which both satisfy the MCL condition (\ref{pole}) and have phase which is stationary with respect to $\mathbf{k}$ will constructively interfere, resulting in a visible wake.

Due to reflection symmetry, we can focus on the $y>0$ half-space without loss of generality. Here, the wake receives contributions from wave vectors with $k_{x,y} > 0$. Then the phase is given by
\begin{equation}
\label{phase_general}
f=\textbf{k}\cdot\textbf{r}=\sqrt{k}\cdot x+\sqrt{k^{2}-k}\cdot y
\end{equation}
where we use Eq.(\ref{pole}) to re-express $k_{x,y}$ in terms of $k$. The stationary phase condition $df/dk = 0$
\begin{equation}
\label{stationary_phase_general}
-\frac{y}{x}=\frac{\sqrt{k-1}}{2k-1}
\end{equation}
is satisfied for $x<0$, which is where the wake is. Equivalent relationships appeared previously~\cite{Lamb,Newman,Faber,Lighthill}. The right-hand side of (\ref{stationary_phase_general}) vanishes at $k=1$ and $k\rightarrow\infty$. It reaches maximum value of $1/2\sqrt{2}$ at $k=3/2$. Thus, Eq.(\ref{stationary_phase_general}) has two solutions for $0\leqslant -y/x < 1/2\sqrt{2}$, corresponding to transverse and diverging wavefronts. These solutions merge at $-y/x=1/2\sqrt{2}$, while none are found above this value. From this we obtain Kelvin's classic result that the wake (for $y>0$) is confined by the angle $\arctan(1/2\sqrt{2}) \approx 19.47^{\circ}$.

Since the phase $f$ is constant along the wavefront, Eqs.(\ref{phase_general}) and (\ref{stationary_phase_general}) can be
solved relative to $x$ and $y$ to give the equation for the wavefront in parametric form:
\begin{equation}
\label{parametric}
x(k)=fk^{-3/2}(2k-1),~~y(k)=-fk^{-3/2}\sqrt{k-1}.
\end{equation}
Wavecrests of the pattern (\ref{parametric}) correspond to the 
following phase choice with integer $n$~\cite{Havelock}
\begin{equation}
\label{wavecrests}
f_{n}=-\pi\left (2n+\frac{5}{4}\right ).
\end{equation}
They are shown in Figure \ref{wakes}a;  the $y<0$ part of the wake is obtained by reflection.  The wake consists of the transverse (t) wavefronts formed by elementary waves with the wave vectors in the $1\leqslant k\leqslant3/2$ range connecting the edges of the pattern across the central line $y=0$, and the diverging (d) wavefronts formed by the waves with the wave vectors in the $k>3/2$ range connecting the source to the edges of the pattern \cite{Kelvin,Lamb,Newman,Faber}.  Combined with the form of the integrand in (\ref{Kelvin_integral}), these facts imply that (for $p(\textbf{k})=const$) the transverse waves are significantly weaker than the diverging ones unless one is close to the central line $y=0$.  The wavelength of the pattern along the latter is $2\pi$ while at the wake boundaries $y/x=\pm1/2\sqrt{2}$ it is $4\pi/3$.  

Alternatively, the wavefront equation can be found 
by substituting solutions of the equation of the stationary phase (\ref{stationary_phase_general}) into the expression for the phase (\ref{phase_general}):
\begin{equation}
\label{phase(x,y)}
f(x,y)=-\frac{x^{2}\left (1+4y^{2}/x^{2}\pm\sqrt{1-8y^{2}/x^{2}}\right )^{3/2}}{4\sqrt{2}y(1\pm\sqrt{1-8y^{2}/x^{2}})}.
\end{equation}
Here the upper and lower signs correspond to the diverging and transverse wavefronts, respectively.  Equating Eqs.(\ref{wavecrests}) and (\ref{phase(x,y)}) then gives two implicit equations for the wavecrests of the transverse and diverging kind.  

The phase function for diverging wavefronts (plus sign in Eq.(\ref{phase(x,y)})) increases in magnitude without bound as $y\rightarrow0$. Here, the water height oscillates with decreasing wavelength, resulting in bunched diverging wavefronts (see Figure~\ref{wakes}a). In practice, factors such as capillarity neglected in the dispersion law (\ref{spectrum}) or non-linear effects can halt this behavior. Possible manifestations are discussed in Refs.~\cite{numerical2,interference}; in practice these effects are limited to a narrow angular range around $y=0$.

\section{Preview}

We contend that the most important physics ingredient that controls gross features of a wake pattern is whether water-piercing is present or not.  A germ of this idea was already stated by He \textit{et al.} \cite{He}. Water-piercing is typical for most ships and it implies rapid variation of the flow near ship's bow and stern.  On the other hand, in the absence of water-piercing the flow variation caused by a disturbance is smooth.  

It is a fundamental theorem of the Fourier analysis that Fourier transform of a smooth and localized function is a smooth and localized function, while Fourier transform of a sharply localized function - the one that is finite within a compact domain and zero elsewhere - is an oscillatory slowly decaying function.  In our case the function in question is the disturbance $\delta p(\textbf{r})$ itself.  

In the absence of water piercing the function $\delta p(\textbf{r})$ is smooth and localized.  Then its Fourier transform $p(\textbf{k})$ that enters the integrands in Eqs.(\ref{general_wake_integral}) and (\ref{Kelvin_integral}) is smooth and localized in the wave vector space.  While this lends support to the idea of the wavelength cutoff having to do with linear size of the source, below we show that its rendition \cite{RM,MR} is incorrect.  Moreover we demonstrate the existence of two types of wake patterns as well as a threshold for the wake formation that previous studies overlooked.

In the presence of water piercing the function $\delta p(\textbf{r})$ has a sharp boundary.  Then its Fourier transform $p(\textbf{k})$ that enters the integrands in Eqs.(\ref{general_wake_integral}) and (\ref{Kelvin_integral}) is a slowly decaying oscillatory function.  While the wavelength corresponding to this periodicity is also related to linear size of the source, it no longer has a meaning of a cutoff.  On the contrary, it is a source of additional interference effects.  Indeed, we show that this line of reasoning naturally leads to a derivation of the interference conditions of Noblesse \textit{et al.} \cite{interference}, thus vindicating their ideas.  Additionally, we provide numerical evidence of the interference effects.   

In order to avoid obscuring physics with details of secondary importance, below we limit ourselves to simple source functions $\delta p(\textbf{r})$ that are characterized by single length scale $a$.  As a result, the only parameter of the problem is the Froude number (\ref{Froude}).  This simplification does not affect generality of our reasoning. 

\section{Effects of finite size and shape:  smooth boundaries}

\subsection {Modified stationary phase argument}

Water-piercing is absent or negligible if the wake is produced by a hovercraft, low-flying airplane, missile, boat in the planing regime \cite{Newman,Faber} or underwater object. This class of sources can be represented by smooth spatially localized source functions $\delta p(\textbf{r})$  whose Fourier transforms $p(\textbf{k})$ are localized in $\textbf{k}$-space. 
This imposes an \textit{ultraviolet} cutoff in the integral (\ref{Kelvin_integral}) which can be approximated by setting a strict upper bound on the magnitude of allowed wave vectors in the integration domain. At this point we observe that this is diametrically opposite to the conjecture of Rabaud and Moisy~\cite{RM} who imposed an \textit{infrared} cutoff that eliminates \textit{large} wavelength contributions. There are no physical grounds for this infrared cutoff, nor for the modification advocated in Ref.~\cite{MR}. 

The size and shape of the now-truncated integration region in Eq.~(\ref{Kelvin_integral}) is encoded in a phenomenological parameter $Q$, which we define as the magnitude of the wave vector which both satisfies the MCL constraint (\ref{pole}) and belongs to the boundary of the now-compact integration domain. Only wave vectors satisfying the following condition
\begin{equation}
\label{range}
k\in [1,Q]
\end{equation}
will contribute to the wake pattern.
Subject to this constraint, the parametric equations for the Kelvin wavefronts (\ref{parametric}) describe those due to sharply localized pressure functions $p(\mathbf{k})$. 

Evolution of these wavefronts for different values of the cutoff parameter $Q$ are shown in Figure~\ref{wakes}. The Kelvin, $Q=\infty$ limit, Figure \ref{wakes}a, differs from the case of finite $Q>3/2$, Figure \ref{wakes}b, in that the diverging wavefronts no longer reach the source at the origin.  Their end points belong to a ray 
\begin{equation}
\label{Froude_ray}
-\frac{y}{x}=\frac{\sqrt{Q-1}}{2Q-1}
\end{equation}
hereafter called the Froude ray.  The latter and its $y<0$ reflection define an inner wake of the opening angle 
\begin{equation}
\label{inner_angle}
2\theta=2\arctan\frac{\sqrt{Q-1}}{2Q-1}\rightarrow \frac{1}{\sqrt{Q}}
\end{equation}   
that only contains the transverse wavefronts;  in the second step we specified to the $Q\gg1$ limit.  We note that both the diverging and transverse wavefronts are found outside the inner wake and are bounded by Kelvin's $39^{\circ}$ angle.  As $Q$ approaches $3/2$, the inner wake widens and the segments of the diverging wavefronts shorten.  At $Q=3/2$ the inner wake angle (\ref{inner_angle}) matches Kelvin's $39^{\circ}$, and the diverging wavefronts disappear.  The wakes having the geometry of Figure \ref{wakes}b have been observed both in nature (see for example, Ref.\cite{Newman}, Figure 6.17, or Ref.\cite{Falkovich}, Figure 3.4) and  numerical studies \cite{DBR,Ellingsen,BDR}.

For $1<Q<3/2$, Figure \ref{wakes}c, the wake consists entirely of the transverse wavefronts confined within the wedge of the angle (\ref{inner_angle}) smaller than Kelvin's.  As the cutoff parameter $Q$ approaches unity, the wake narrows and disappears for $Q<1$.  The absence of the wake for $Q<1$ does not imply strict lack of the waves but indicates that the stationary phase approximation no longer applies, and that the waves are weak.  

The wakes generated by finite-size pressure sources, like those in Figures \ref{wakes}b and \ref{wakes}c, are hereafter called Kelvin-Froude wakes.  

\subsection{Sharp wave vector cutoff}

To illustrate the effects of this ultraviolet cutoff, we now consider an example where the integration domain in Eq.(\ref{Kelvin_integral}) is a $2k_{max} \times 2k_{max}$ square centered at the origin as shown in Figure~\ref{MCL}. The intersections of the MCL curve (\ref{pole}) with the boundary of this domain determine the cutoff parameter $Q$:
\begin{equation}
\label{squareQ}
  Q = \left \{
\begin{aligned}
    &k_{max}^{2}, && \text{if}\ k_{max}^{2}<2 \\
    &\frac{1+\sqrt{1+4k_{max}^{2}}}{2}, && \text{if}\ k_{max}^{2}\geqslant2
\end{aligned} \right.
\end{equation} 
There are three possibilities. For $k_{max} < 1$ (brown square) there will be no wake. For $1 \leq k_{max} \leq (3/2)^{1/2}$ (purple square) the wake will have the geometry of Figure~\ref{wakes}c. For $k_{max} > (3/2)^{1/2}$ (blue square) the wake will have the geometry of Figure~\ref{wakes}b. The wavecrests will be given by Eqs.(\ref{parametric}), (\ref{wavecrests}), (\ref{range}), and (\ref{squareQ}).

\begin{figure*}[!t]
\centering
\includegraphics[width=2.0\columnwidth, keepaspectratio]{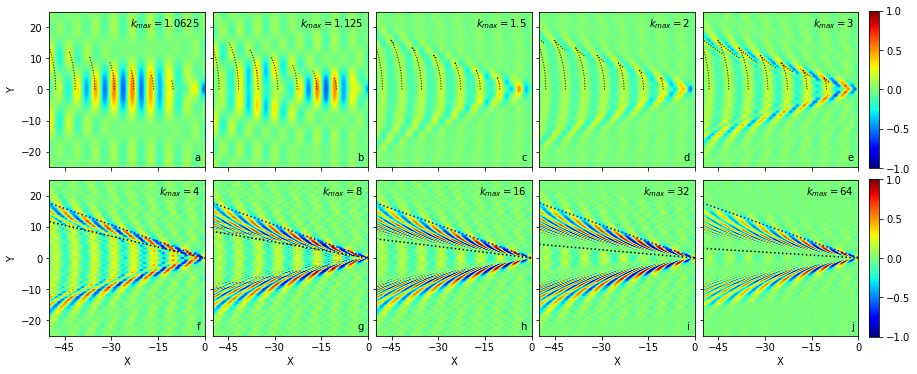} 
\caption{(Color online) Evolution of the wake pattern generated  by Eq.(\ref{Kelvin_integral}) for the square integration domain $|k_{x,y}|\leqslant k_{max}$ for a series of different values of $k_{max}$ shown as color-coded topographical map of the water displacement $\zeta(\textbf{r})$ rescaled to fit on the $[-1, 1]$ range as indicated by the color bars at the far right of each row of images.  In the top row, cases (a)-(e), black-dotted curves represent the wavecrests given by Eqs.(\ref{parametric}), (\ref{wavecrests}), (\ref{range}) and (\ref{squareQ}).  In the bottom row, cases (f)-(j), the black- and blue-dotted straight lines represent the Froude (\ref{Froude_ray}) and Kelvin, $-y/x=1/2\sqrt{2}$, rays, respectively.}
\label{sharp}
\end{figure*}

We computed the integral (\ref{Kelvin_integral}) numerically for several different value of $k_{max}$. Relief plots of the water displacement $\zeta(\mathbf{r})$ are shown in Figure~\ref{sharp}. These results fully support phenomenological reasoning based on the modified stationary phase argument.

In the top row of images (a-e), we overlay the curves given by Eqs.(\ref{parametric}), (\ref{wavecrests}), (\ref{range}), and (\ref{squareQ}) for $y>0$. For $k_{max}<3/2$ (a-b) the diverging wavefronts are absent and the wakes are narrow and weak. As $k_{max}$ increases beyond $3/2$ (c-e) the diverging wavefronts grow in presence and magnitude.  

In the bottom row of images (f-j), we overlay the Kelvin ($-y/x=1/2\sqrt{2}$) and Froude (\ref{Froude_ray}) rays. These wakes exhibit well-developed diverging wavefronts and a narrowing inner wake with increasing $k_{max}$, as predicted by Eqs.(\ref{inner_angle}) and (\ref{squareQ}). In the final three images (h-j) there is a sliver of nearly zero water displacement between the Froude ray and the outer wake containing both diverging and transverse wavefronts. This is an artifact reflecting our inability to display rapidly oscillating diverging wavefronts, a property we discussed following Eq.(\ref{phase(x,y)}).  We then conclude that as $k_{max}\rightarrow \infty$, the wake pattern approaches Kelvin's ideal limit sketched in Figure \ref{wakes}a.  

Since $k_{max}$ corresponds to an inverse spatial scale of the pressure source, these conclusions can be recast in terms of the Froude number $F\equiv \sqrt{k_{max}}$.
There exist two critical Froude numbers, $F_{1}=1$ and $F_{2}=(3/2)^{1/4}$, such 
that no wake is found for $F< F_{1}$, for $F_{1}\leqslant F\leqslant F_{2}$ the wake has the geometry of Figure \ref{wakes}c, while for $F>F_{2}$ the wake has the geometry of Figure \ref{wakes}b.  

A different implementation of the sharp ultraviolet cutoff in the integral (\ref{Kelvin_integral}) only changes the critical values $F_{1,2}$ without affecting the gross picture that there are two possibilities for the wake patterns controlled by the Froude number and encapsulated in Figures \ref{wakes}b or \ref{wakes}c.  
The underlying physics is a suppression of the short-wavelength modes by finite size effects which, as $F$ decreases, progressively diminishes diverging wavefronts until they disappear at $F=F_2$ and then gradually excludes 
the transverse wavefronts until the wake is gone at $F=F_{1}$. 
This quantitative picture substantiates and extends qualitative reasoning regarding the dependence of the wake pattern on the Froude number \cite{Lighthill}.  

The existence of the lower critical Froude number $F_{1}$ is consistent with interpretation of the Froude number (\ref{Froude}) as a Mach number (\ref{Mach}).

\subsection{Generalization to smooth pressure sources $p(\textbf{k})$}

While the wakes of sources sharply localized in the wave vector space can be understood by generalization of the argument that explains the ideal Kelvin wake pattern, they mimic realistic disturbances poorly.
Their direct space counterparts $\delta p(\textbf{r})$ feature non-physical slowly decaying oscillatory behavior as one moves away from the pressure center.  Yet, we contend that the reasoning developed to understand these cases largely carries over to smooth localized pressure disturbances $p(\textbf{k})$ whose real space counterparts $\delta p(\textbf{r})$ are free of artifacts.  

Indeed, the origin of the critical Froude numbers $F_{1,2}$ can be traced back to the existence of the two critical values of the cutoff parameter (\ref{range}).  The first, $Q_1=1$, sets the threshold for the wake appearing in the form of transverse wavefronts. The second, $Q_2=3/2$ determines the onset of diverging wavefronts.
However, for smooth $p(\textbf{k})$ the cutoff parameter $Q$ is no longer sharply defined because contribution of large wave vector modes into the wake pattern is only suppressed rather than excluded.  The well-localized nature of the pressure function $p(\textbf{k})$ still allows us to define a characteristic cutoff parameter $Q$.  The instants when $Q$ reaches the threshold values of $Q_{1}=1$ and  $Q_{2}=3/2$ define the characteristic Froude numbers $F_{1}$ and $F_{2}$, respectively.  Thus the transitions at $F=F_{1,2}$ become smeared.  Moreover, observation of the narrow wake pattern like in Figure \ref{wakes}c may become problematic as the tail of the function $p(\textbf{k})$ admits high wave vector modes, causing the wake to appear wider.   

To test these predictions, we focus on smooth pressure distributions characterized by a single length scale $a\equiv 1/F^{2}$ (see Eq.(\ref{Froude})).  Then the response to the moving pressure source is given by a version of Eq.(\ref{Kelvin_integral})       
\begin{equation}
\label{beyond_Kelvin_integral}
\zeta(\textbf{r})=\int\frac{d^{2}k}{(2\pi)^{2}}\frac{kp(\textbf{k}/F^{2})e^{i\textbf{k}\cdot\textbf{r}}}{(k_{x}+i0)^{2}-k}
\end{equation}
with dependence on the Froude number explicitly displayed.  For isotropic disturbance $p(\textbf{k}/F^{2})\equiv p(k/F^{2})$ the cutoff parameter $Q$ (\ref{range}) then can be estimated as  
\begin{equation}
\label{isotropic_cutoff}
Q\simeq F^{2}
\end{equation}
As a result the opening angle of the inner wake (\ref{inner_angle}) for $F\gg1$ behaves as
\begin{equation}
\label{limit_iso_angle}
2\theta \simeq \frac{1}{F}
\end{equation}
The $1/F$ asymptotic behavior was predicted for the angle of largest waves due to isotropic Gaussian pressure source \cite{RM,DBR}.  The result (\ref{limit_iso_angle}) makes apparent a parallel between the Froude (\ref{Froude}) and Mach (\ref{Mach}) numbers but limits quantitative scope of the analogy \cite{Falkovich} to isotropic pressure sources as the next example shows.    
 
For strongly anisotropic disturbance $p(\textbf{k}/F^{2})\equiv p(k_{x}/F^{2})$ the ${x}$-components of the wave vectors are effectively suppressed at $k_{x}\simeq F^{2}$;  the MCL condition (\ref{pole}) then implies that 
\begin{equation}
\label{anisotropic_cutoff}
Q\simeq F^{4}
\end{equation}         
In this case the opening angle of the inner wake (\ref{inner_angle}) for $F\gg1$ behaves as 
\begin{equation}
\label{limit_aniso_angle}
2\theta \simeq \frac{1}{F^{2}}
\end{equation}
which is not a Mach-like behavior.  This $1/F^2$ asymptotic behavior was predicted by Refs.~\cite{interference,MR} for two distinct model anisotropic pressure disturbances.

Below we evaluate the integral (\ref{beyond_Kelvin_integral}) numerically for two distinct isotropic and strongly anisotropic pressure disturbances to which the predictions (\ref{isotropic_cutoff})-(\ref{limit_aniso_angle}) apply. Here, we ensure that $F^2\ll k_{max}$ for the isotropic case and $F^4 \ll k_{max}$ for the strongly anisotropic one, to probe effects of the pressure source rather than the cutoff.

\subsubsection{Isotropic Gaussian pressure source}
  
As an example of an isotropic pressure source we look at the Gaussian disturbance
\begin{equation}
\label{iGaussian_disturbance}
p\left (\frac{\textbf{k}}{F^{2}}\right )=\exp\left (-\frac{k^{2}}{F^{4}}\right ).
\end{equation} 
While this case has been studied previously \cite{DBR,Ellingsen}, our focus is different as we test quantitative reasoning regarding evolution of the wake pattern with the Froude number.  For low $F$, we disagree with Refs.\cite{DBR,Ellingsen}.  
The analysis presented in Refs. \cite{DBR,Ellingsen} is based on the far-field approximation (\ref{far-field}) which is not accurate for $F$ small.  
Our convention (\ref{iGaussian_disturbance}) also gives a Froude number 
$\sqrt{2\pi}$ times larger than that %employed 
in Refs.\cite{DBR,Ellingsen}.

We computed the integral (\ref{beyond_Kelvin_integral}) numerically for this pressure source using different values of $F$. 
\begin{figure*}[!t]
\centering
\includegraphics[width=2.0\columnwidth, keepaspectratio]{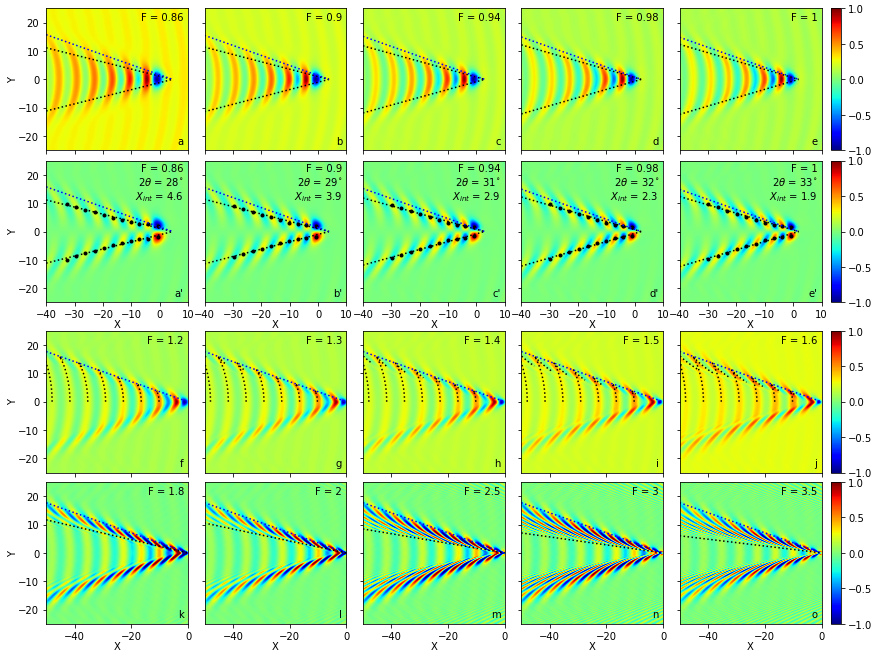} 
\caption{Evolution of the wake pattern generated  by isotropic Gaussian pressure source, Eqs.(\ref{beyond_Kelvin_integral}) and (\ref{iGaussian_disturbance}), for a series of different values of the Froude number $F$.  The legend is the same as that of Figure \ref{sharp}.  The bounding rays for the top row were drawn via a procedure based on the topographical map of $\partial \zeta/\partial y$, the second from the top row, and explained in the main text.   The source point was set as the intersection of these bounding rays with the $y=0$ line, and was found to be at some $x=X_{int}>0$ for $F<F_{2}$.   The cutoff parameter employed in the third from the top row for the wavecrests (\ref{parametric}) and in the fourth from the top row for the Froude ray (\ref{Froude_ray}) is $Q=1.4F^{2}$. }
\label{isogaussian}
\end{figure*} 
The results displayed in Figure \ref{isogaussian} largely support the reasoning regarding the existence of two characteristic Froude numbers $F_{1,2}$ and two types of the wake patterns having the geometries of Figures \ref{wakes}b and \ref{wakes}c.   Specifically, the results are reasonably well-described in terms of the cutoff parameter $Q\approx 1.4 F^{2}$ (in agreement with the scaling prediction (\ref{isotropic_cutoff})) and characteristic Froude numbers $F_{1} \approx 0.84$ and $F_{2} \approx 1.04$.  

The wakes in the first row of images (a-e) feature only transverse wavefronts confined by a wedge with bounding angle smaller than Kelvin's. We obtained the bounding wedge by fitting lines through points of maximal $|\partial \zeta / \partial y|$ as demonstrated in the second row (a'-e'). The Kelvin ray ($-y/x=1/2\sqrt{2}$) is included for comparison.  As the Froude number increases 
toward $F=F_{2}$, the wake widens approaching the Kelvin $39^{\circ}$ limit.  
This would agree with the geometry of Figure~\ref{wakes}c, except that the apex of the wake appears \textit{ahead} of the source at $x=X_{int}>0$. This feature is commonly seen in marine practice (see, for example, Figure 6.17 of Ref.\cite{Newman}).  As the Froude number increases toward $F=F_{2}$, the apex
approaches the disturbance center $x=0$. Numerically determined values of the wake angle $2\theta$ and the apex position $X_{int}$ for each $F$ are quoted in images 
(a'-e'). In this range of low Froude numbers, we disagree with Refs.~\cite{DBR,Ellingsen} which found that the wake is always bounded
by Kelvin's $39^{\circ}$ angle.

In the third row (f-j), we overlay the curves
given by Eqs. (\ref{parametric}), (\ref{wavecrests}), and (\ref{range}) with $Q=1.4 F^{2}$.  
As the Froude number increases from $1.2$ to $1.6$, the diverging wavefronts develop and grow in presence.  In the bottom row (k-o), we overlay the Kelvin $-y/x=1/2\sqrt{2}$ and Froude (\ref{Froude_ray}) rays. These clearly demonstrate well-developed diverging wavefronts and narrowing of the inner wake with
increasing $F$ in accordance with Eq.(\ref{inner_angle}). Upon visual comparison, the wakes in the range $F=1.25-2.5$ agree with those in Refs.~\cite{DBR,Ellingsen} in both appearance and evolution with Froude number.

Disturbance due to the small submerged ball is closely related to the isotropic Gaussian pressure distribution.  Indeed, the effective pressure distribution in this case (\ref{ball}) is only marginally anisotropic, implying that appearance of the wake pattern will be dominated by the isotropic exponential factor in (\ref{ball}).   Therefore one expects that the wake patterns evolve with the (depth-based) Froude number (\ref{Froude}) 
in a manner qualitatively similar to the isotropic Gaussian pressure source.  We calculated the wake integral (\ref{beyond_Kelvin_integral}) with the effective pressure distribution (\ref{ball}) and confirmed  this expectation which also includes existence of small-$F$ narrow wakes with apex ahead of the source.  We chose not to display corresponding results as the difference from the isotropic Gaussian case is purely quantitative, specifically, characteristic Froude numbers $F_{1,2}$ are different.   Our results overlap with earlier numerical calculation limited to a large Froude number regime \cite{Peti} and with more recent evaluation \cite{AK} that covered a broader range of the Froude numbers.  

\subsubsection{Strongly anisotropic Gaussian pressure source}

\begin{figure*}[!t]
\centering
\includegraphics[width=2.0\columnwidth]{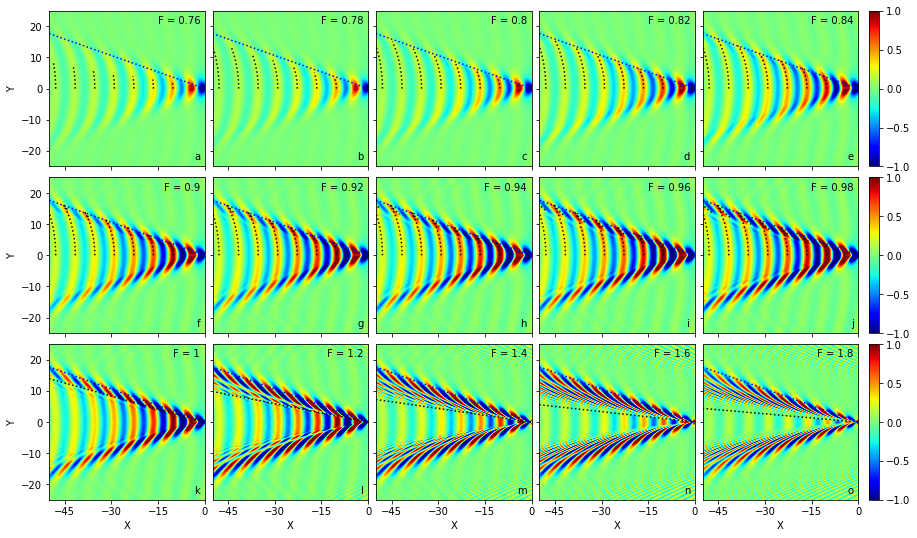}
\caption{Evolution of the wake pattern generated  by strongly anisotropic Gaussian pressure source, Eqs.(\ref{beyond_Kelvin_integral}) and (\ref{aGaussian_disturbance}), for a series of different values of the Froude number $F$.  The legend is the same as that of Figures \ref{sharp} and \ref{isogaussian}. The cutoff parameter employed in the first and second rows for the wavecrests, Eqs.(\ref{parametric}) and (\ref{wavecrests}), and in the third row for the Froude ray (\ref{Froude_ray}) is $Q=3.1F^{4}$.}
\label{anisogaussian}
\end{figure*}

As an example of a strongly anisotropic pressure source we look at the Gaussian disturbance
\begin{equation}
\label{aGaussian_disturbance}
p\left (\frac{\textbf{k}}{F^{2}}\right )=\exp\left (-\frac{k_{x}^{2}}{F^{4}}\right ).
\end{equation} 
We numerically evaluated the integral (\ref{beyond_Kelvin_integral}) for a range of $F$. The results, shown in Figure~\ref{anisogaussian}, largely support the existence of two characteristic Froude numbers $F_{1,2}$ and the two types of wake patterns shown in Figure~\ref{wakes}b,c.
Specifically, we find the cutoff parameter $Q\approx 3.1F^{4}$ (in agreement with the scaling prediction (\ref{anisotropic_cutoff})) and characteristic Froude numbers $F_{1} \approx 0.75$ and $F_{2} \approx 0.83$.  

The wakes in the first row images, (a)-(e), only feature transverse wavefronts bounded by the Kelvin ray $-y/x=1/2\sqrt{2}$. This would be an illustration of the geometry of Figure \ref{wakes}c if not for the fact that the wake is wider (a possibility already anticipated);  the wake angle appears to be exactly Kelvin's $39^{\circ}$.   In contrast to the low-$F$ regime of isotropic case, Figure \ref{isogaussian} (a-e), the apex of the wake coincides with the center of the source.  It is curious that low-$F$ wakes of isotropic pressure source, Figure \ref{isogaussian},  are more \textit{narrow} than low-$F$ wakes of strongly anisotropic one, Figure \ref{anisogaussian}.   

In the first and second row of images (a-j), we overlay the curves 
given by Eqs. (\ref{parametric}), (\ref{wavecrests}), and (\ref{range}) with $Q=3.1F^{4}$.  As the Froude number increases from $0.9$ to $0.98$, the diverging wavefronts develop and grow in presence.  

In the bottom row (k-o), we overlay the Kelvin $-y/x=1/2\sqrt{2}$ and Froude (\ref{Froude_ray}) rays. These wakes clearly demonstrate well-developed diverging wavefronts and narrowing of the inner wake with increasing $F$ in accordance with Eq.(\ref{Froude_ray}).  

\section{Effects of finite size and shape:  sharp boundaries}

Due to water piercing, effective pressure disturbances describing actual ships correspond to functions $\delta p(\textbf{r})$ that are finite within a compact spatial domain and zero otherwise.  Their Fourier transforms $p(\textbf{k})$ entering Eq.(\ref{general_wake_integral}) are slowly decaying oscillating functions.
As they do not suppress large wave vector modes, the wake is always present. The oscillatory behavior of $p(\mathbf{k})$, which introduces additional interference effects, must be treated on equal footing as the oscillating exponential in (\ref{general_wake_integral}).

\subsection{Finite length segment constant pressure source: two-point interference argument}

Rather than attempt to discuss all possible sources, we will focus on the case of a boundary with two sharp corners.  Specifically, we will be employing the function
\begin{equation}
\label{segment}
p\left (\frac{\textbf{k}}{F^{2}}\right )=\frac{2F^{2}}{k_{x}}\sin\left (\frac{k_{x}}{2F^{2}}\right )\equiv\frac{2}{k_{x}a}\sin\left (\frac{k_{x}a}{2}\right )
\end{equation}
which corresponds to a constant pressure line segment of length $a\equiv 1/F^{2}$ parallel to the $x$-axis and centered at the origin.  The function (\ref{segment}) mimics the geometry of a slender ship.  Substituting (\ref{segment}) into Eq.(\ref{beyond_Kelvin_integral}) we find
\begin{eqnarray}
\label{segment_integral}
\zeta(\textbf{r})&=&\int\frac{2F^{2}kd^{2}k}{(2\pi)^{2}k_{x}}\frac{\sin(k_{x}/2F^{2})e^{i\textbf{k}\cdot\textbf{r}}}{(k_{x}+i0)^{2}-k}\nonumber\\
&=&\int\frac{kd^{2}k}{(2\pi)^{2}ik_{x}a}\frac{e^{i[k_{x}(x+a/2)+k_{y}y]}-e^{i[k_{x}(x-a/2)+k_{y}y]}}{(k_{x}+i0)^{2}-k}\nonumber\\
\end{eqnarray}
A classic idea in marine hydrodynamics proposes that a ship wake can be understood as the interference of two opposite-effect Kelvin wakes from the bow and stern~\cite{resistance,Newman}. The second representation in (\ref{segment_integral}) supports this interpretation, with two superimposed wakes emanating from the segment ends $x=\mp a/2$. However, the two wakes in Eq.(\ref{segment_integral}) are not exactly Kelvin's and the function (\ref{segment}) is not a real-space pressure dipole.  This distinction is important because the function $p(\textbf{k}/F^{2})$ in Eq.(\ref{beyond_Kelvin_integral}) is the property of the pressure source at rest.  Thus if a vessel is symmetric about its center, so is the pressure function;  a pressure dipole is incompatible with central symmetry.    

In the stationary phase approximation, the wakes originating at 
$x=\mp a/2$ are described by 
phase functions $f(x\pm a/2,y)$ 
of the form (\ref{phase(x,y)}) found in the Kelvin case.  
The two wakes interfere for $x<-a/2$ and $|y/(x+a/2)|<1/2\sqrt{2}$, while the rest of the pattern is caused by he Kelvin-like wake generated by the segment front $x=a/2$. Our main focus is the wedge-shaped region of two-point interference.  

The locus of the points of largest waves is given by the condition of constructive interference
 \begin{equation}
\label{constructive}
f\left (x+\frac{a}{2}\right )-f\left (x-\frac{a}{2}\right )=\pi (2n+1), ~~~n=0, 1, 2,...
\end{equation}
Likewise, the locus of the points of smallest waves is given by a similar expression with $2n+1$ replaced by $2n$.  The relationship (\ref{constructive}) encompasses two different equations corresponding to the two sign choices in the phase function (\ref{phase(x,y)}). The upper and lower signs correspond to interfering diverging (dd) wavefronts and transverse (tt) wavefronts, respectively. We will use the abbreviations dd or tt to refer to the two-point interference effects. 

Far from the source, $|x|\gg a$,  Eq.(\ref{constructive}) simplifies to  
\begin{equation}
\label{same_constructive}
\frac{\partial f}{\partial x}=\frac{\pi(2n+1)}{a}\equiv\pi(2n+1)F^{2}.
\end{equation}  
Curiously, evaluation of the derivative $\partial f/\partial x$ with two different phase functions (\ref{phase(x,y)}) leads to the same result 
\begin{equation}
\label{discrete_stationary_phase}
-\frac{y}{x}=\frac{\sqrt{k_{n}-1}}{2k_{n}-1},~~~k_{n}=\pi^{2}(2n+1)^{2}F^{4}
\end{equation}  
Here, tt and dd interference take place in the regions $1 \leq k_n \leq 3/2$ and $k_n > 3/2$, respectively. 
Eq.(\ref{discrete_stationary_phase}) is a discrete counterpart of the 
stationary phase condition (\ref{stationary_phase_general}).  The wave vector magnitude $k$ is replaced by the discrete set of $k_{n}$ hereafter called the wave numbers.

Ref.~\cite{interference} obtained the relationship~(\ref{discrete_stationary_phase}) by assuming the wake pattern is caused by two interfering opposite-effect Kelvin wakes off the bow and stern, mimicked by a pressure dipole. Our analysis vindicates this reasoning (we also derived the corresponding interference condition for a catamaran~\cite{interference}). The physical picture of two interfering wakes arises naturally from the integral (\ref{segment_integral}), which also gives insight into the magnitude of the effects. Hereafter our analysis complements that of Ref.\cite{interference}; numerically illustration of these interference effects is where our results largely lie.

Subjecting the wave numbers $k_n$ (\ref{discrete_stationary_phase}) to the MCL condition (\ref{pole}) gives a set of wave vector components, 
$k_{x,n}=\pi(2n+1)F^{2}=\pi(2n+1)/a\geqslant \pi/a$ familiar from analysis of a classic one-dimensional toy model allowing only transverse wavefronts \cite{Newman}.   

Constructive tt interference occurs on the central line $y=0$ for a discrete set of Froude numbers, which satisfy $k_n=1$ in Eq.~(\ref{discrete_stationary_phase}).
\begin{equation}
\label{y=0_constructive_interference}
F_{n}=\frac{1}{\sqrt{\pi(2n+1)}}\approx 0.564, 0.326, 0.252, 0.213,...
\end{equation}
Likewise, destructive tt inteference occurs for
\begin{equation}
\label{y=0_destructive_interference}
F_{n}=\frac{1}{\sqrt{2\pi n}}\approx 0.399,0.282,0.230,0.199,...
\end{equation}
These two sets are familiar from analysis of the aforementioned 
toy model admitting only transverse wavefronts \cite{Newman}. 
We recover this model along the central line where $f(x,0)=x$, Eq.(\ref{phase(x,y)}). The "unfavorable" set (\ref{y=0_constructive_interference}) leading to large waves  
corresponds to maxima of the wave resistance versus Froude number dependence \cite{Newman}.  For the  "favorable" set (\ref{y=0_destructive_interference}) the waves are small, and the wave resistance exhibits minima \cite{Newman}.

\begin{figure*}[t!]
\centering
\includegraphics[width=2.0\columnwidth, keepaspectratio]{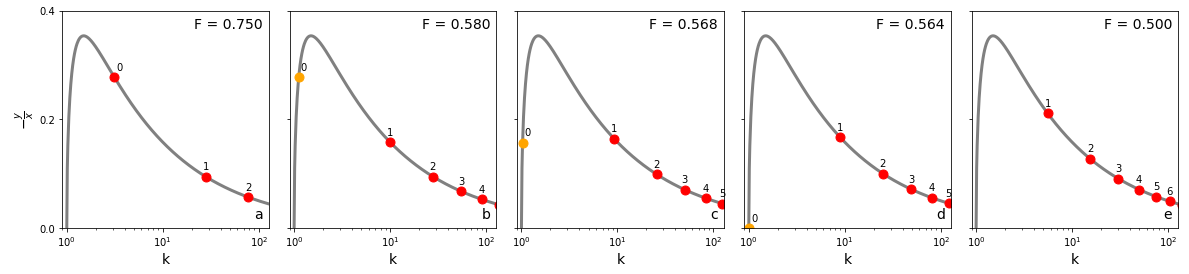} 
\caption{(Color online) Flow of the wave numbers $k_{n}$ (\ref{discrete_stationary_phase}) shown as solid red (dd interference) or orange (tt interference) circles visible within the $1\leqslant k \leqslant 128$ range along the curve of the stationary phase (\ref{stationary_phase_general}) (opaque grayscale) presented in the semi-logarithmic scale with decrease of the Froude number $F$.  Numbers next to the circles are subscripts $n$ in $k_{n}$ labeling constructive interference fringes.}
\label{flow}
\end{figure*}

The "flow" of wave numbers $k_n$ (\ref{discrete_stationary_phase}) along the curve of stationary phase (\ref{stationary_phase_general}), shown in Figure~\ref{flow}, encodes how the wake evolves with \textit{decreasing} Froude number.
Here, we plot the stationary phase curve (\ref{stationary_phase_general}) in grayscale and denote wave numbers $k_n$ visible in the range $1 \leqslant k \leqslant k_{max}$ with colored circles, which we annotate with their index $n$. Red and orange circles correspond to dd and tt interference effects, respectively.

If $\pi^{2}F^{4}>3/2$ or $F>0.624$ all the wave numbers belong to the descending, $k>3/2$, part of the curve and represent the dd interference effects (Figure~\ref{flow}a).  The entire interference pattern is confined within a wedge of an angle given by Eq.(\ref{inner_angle}) with 
\begin{equation}
\label{zero_Q}
Q=k_{0}=\pi^{2}F^{4}
\end{equation}

As $F$ decreases, the wave numbers flow leftward along the stationary phase curve. The confining angle, given by Eqs.(\ref{inner_angle}) and (\ref{zero_Q}) approaches $39^{\circ}$ as $F\rightarrow (3/2\pi^2)^{1/4}=0.624$ from above. At smaller $F$, the wave number $k_0$ shifts to the ascending, $k<3/2$, segment of the curve (\ref{stationary_phase_general}) and represents tt interference. At the same time the remaining wave numbers, $k_{n}$, $n\geqslant 1$, still belong to the $k>3/2$ segment. The confining angle now decreases with decreasing $F$ as seen in Figure \ref{flow}b.

As the Froude number further decreases, $k_{0}$ moves down while the remaining $k_{n}$'s move up the curve (\ref{stationary_phase_general}).  The confining angle 
decreases and is still given by Eqs.(\ref{inner_angle}) and (\ref{zero_Q}) until $k_{0}$ and $k_{1}$ reach the same height. 
This point, $F=0.568$ (Figure~\ref{flow}c), is very close to the $F_0=0.564$, Eq.(\ref{y=0_constructive_interference}), threshold. Thus, the interference pattern is confined within a narrow wedge.

At even lower $F$, the entire interference pattern is confined by a wedge with angle given by Eq.(\ref{inner_angle}) with
\begin{equation}
\label{one_Q}
Q=k_{1}=9\pi^{2}F^{4}
\end{equation}  
It contains the single $k_{0}$ tt interference fringe and an infinite set of the $n\geqslant1$, dd interference fringes.   As the Froude number further decreases toward $F=F_{0}=0.564$, the $k_{0}$ tt fringe approaches the central line $y=0$
and the confining wedge widens. At $F=0.654$ (Figure~\ref{flow}d), the tt fringe is nearly extinct.

For $F$ slightly smaller than $0.564$ (Figure~\ref{flow}e), only the $n\geqslant1$ dd interference fringes are present. While this looks similar to the case in Figure~\ref{flow}a with the substitution $n\rightarrow n+1$, this is only a half-cycle. There are also "destructive" ($2n+1\rightarrow2n$) wave numbers sandwiched between neighboring "constructive" $k_n$'s (\ref{discrete_stationary_phase}). To complete the cycle, the Froude number must further decrease until the $k_1$ tt fringe is extinct.

\subsection{Numerical visibility requirements}

We numerically integrated (\ref{segment_integral}) over a $2k_{max}\times 2k_{max}$ domain with $k_{max}=128\gg1$. To properly understand numerical results, we note that the wakes generated by the segment's ends have the geometry of Figure~\ref{wakes}b.  They feature inner wakes with opening angle (\ref{inner_angle})     
\begin{equation}
\label{inner_angle_segment}
2\theta\approx\frac{1}{\sqrt{Q}}\approx\frac{1}{\sqrt{k_{max}}}=\frac{1}{8\sqrt{2}}
\end{equation}
This is a manifestation of strict elimination of the diverging $k>k_{max}$ wave modes
, which also limits the wave numbers $k_n$ (\ref{discrete_stationary_phase}) participating in constructive interference to those satisfying $k_n<k_{max}$.
Equating $k_{n}$ and $k_{max}$ gives the "constructive" Froude numbers
\begin{equation}
\label{constructive_F}
F_{c}^{(n)}=\frac{k_{max}^{1/4}}{\sqrt{\pi (2n+1)}}=1.898, 1.096, 0.849, 0.717,...
\end{equation}   
Similarly, the "destructive" ($2n+1\rightarrow 2n$) case gives
\begin{equation}
\label{destructive_F}
F_{d}^{(n)}=\frac{k_{max}^{1/4}}{\sqrt{2\pi n}}=1.342, 0.949, 0.775, 0.671...
\end{equation}   
For a given Froude number $F$, only the interference fringes whose order $n$ and type satisfy the inequalities $F<F_{c,d}^{(n)}$ are visible. For $k_{max}=128$ the two-point interference effects are only visible  if $F<F_{c}^{(0)}=1.898$.  Figure~\ref{flow} shows the visible fringes for several $F$, as the horizontal axis covers the range $1\leqslant k\leqslant 128$. The destructive fringes sandwiched between neighboring constructive ones are not shown.

\subsection{Numerical confirmation}

\begin{figure*}[t!]
\centering
\includegraphics[width=2.0\columnwidth, keepaspectratio]{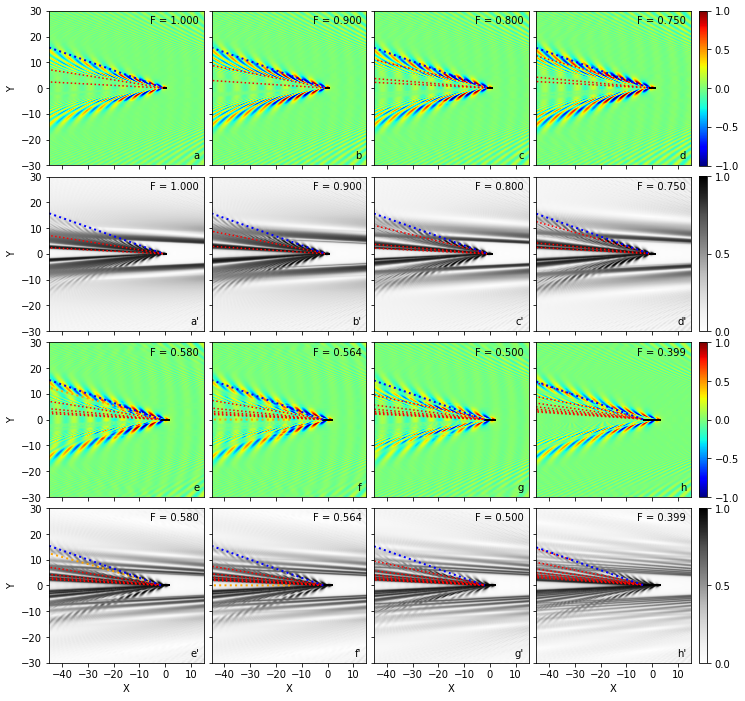} 
\caption{(Color online) Evolution of the two-point interference effects in the wake pattern generated  by the uniform pressure segment source of length $a=1/F^{2}$, Eq.(\ref{segment_integral}), for a series of moderately large Froude numbers $F$.  The first and third rows display topographical map of $\zeta(\textbf{r})$ while its magnitude $|\zeta(\textbf{r})|$ (grayscale) is presented in the second and fourth rows followed by grayscale bars.   The source is shown as a bold segment centered at the origin, and each of the wakes is overlaid with the Kelvin ray $-y/(x+a/2) =1/2\sqrt{2}$  (blue dotted line).  The loci of the points of constructive interference given by Eqs.(\ref{constructive}) and (\ref{phase(x,y)}) for all visible $n$ are indicated by red dotted lines if they represent the dd interference effects, $k_{n}>3/2$, or orange dotted lines if the effect has the tt nature, $1\leqslant k_{n}\leqslant3/2$.}
\label{1stcycle}
\end{figure*}

In Figure~\ref{1stcycle}, we show numerically computed wake patterns for Froude numbers $F = 1.00-0.399$.
The pressure segment sourcing the wake is shown as a bold interval of length $a=1/F^{2}$ centered at the origin.  We overlay the Kelvin ray $-y/(x+a/2)=1/2\sqrt{2}$ originating from the segment end $x=a/2$ with a blue dotted line. Red and orange dotted lines denote the loci of points of constructive interference (given by Eqs.(\ref{constructive}) and (\ref{phase(x,y)})) for all visible $n$ for dd ($k_n>3/2$) and tt ($1\leqslant k_n \leqslant 3/2$) effects, respectively. While for larger $F$ these are nearly identical to the straight line rays (\ref{discrete_stationary_phase}), the more accurate expressions (\ref{constructive}) and (\ref{phase(x,y)}) make a difference at smaller $F$.

Because $k_{max}=128$ is sufficiently large, our inability to properly display rapidly oscillating functions comes into play before the wave vector cutoff effect. As a result, dd interference effects are invisible in plots of $\zeta(\mathbf{r})$. Thus, the inner wakes appear much wider than the prediction $2\theta=1/8\sqrt{2}$.  
We have seen this artifact before for the point source (Figures~\ref{sharp}(i, j)), where there appeared to be a sliver of nearly zero water displacement %sandwiched 
between the Froude ray (\ref{Froude_ray}) and the part of the wake containing both the diverging and transverse wavefronts.  

To reveal the dd interference effects, we include plots of the wake magnitude $|\zeta(\mathbf{r})|$ in Figure~\ref{1stcycle} below each plot of $\zeta$. Here, rapid oscillations in $|\zeta|$ average to a positive constant, creating a smooth background upon which interference effects are visible. Constructive fringes appear dark while destructive fringes appear light. Each magnitude plot features a  
narrow innermost wake, the finite $k_{max}$ effect, whose opening angle is indeed given by $2\theta=1/2\sqrt{8}$.   Evolution of the $|\zeta|$ pattern with the Froude number fully agrees with our analysis. To aid in counting fringes, we chose Froude numbers in Figure~\ref{1stcycle}(d'-g') to coincide with those in Figure \ref{flow} (a, b, d, and e). 

As $F$ decreases,  dark dd fringes get progressively wider as their angle increases toward Kelvin's. 
New narrow fringes become visible at the threshold Froude numbers satisfying the condition $k_{n}=k_{max}$.  
The maps of $|\zeta|$ also reveal "echoes" which represent magnified patterns of $|\zeta|$ in the vicinity of the inner wake boundaries.  These artifacts of finite $k_{max}$ make it easier to discern a large number of fringes in the case of $F$ small. 

At $F=0.580$ the $n=0$ fringe ends up on the transverse part of the curve of the stationary phase (Figure~\ref{flow}b), and we do not see tt interference effects in $|\zeta|$.  At the same time, the magnitude of the transverse waves in the original wake pattern $\zeta(\mathbf{r})$ steadily increases as $F$ decreases in the range $1.000-0.564$.
The lower boundary of this interval corresponds to the first unfavorable Froude number $F_{0}$ (\ref{y=0_constructive_interference});  the $n=0$ tt fringe which is now along the central line $y=0$ is about to disappear.  The transverse waves are indeed largest at $F=F_{0}$ but this maximum is flat and the effect is weak.  
This is visible from Figure~\ref{1stcycle}(e-g) which feature transverse waves of comparable magnitude.  On the other hand, Figure~\ref{1stcycle}h, corresponding to the first favorable Froude number $F=F_1=0.399$ (\ref{y=0_destructive_interference}), exhibits strong destructive interference with nearly invisible transverse wavefronts.

\begin{figure*}[!t]
\centering
\includegraphics[width=2.0\columnwidth, keepaspectratio]{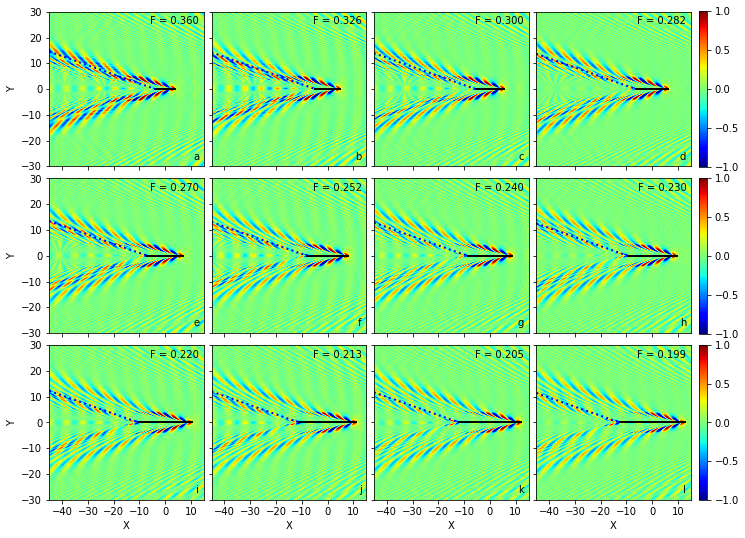} 
\caption{(Color online) Evolution of the wake pattern generated  by the uniform pressure segment source of length $a=1/F^{2}$, Eq.(\ref{segment_integral}), for a series of small Froude numbers $F$.  Each of the wakes is overlaid with the Kelvin ray $-y/(x+a/2) =1/2\sqrt{2}$ (blue dotted line).}
\label{_cycles}
\end{figure*}

While we have confirmed the two-point interference picture of wake patterns, it is not obvious from Figure~\ref{1stcycle} that we are dealing with two interfering wakes. To see this, we must look at even smaller Froude numbers where the pressure source is sufficiently long and the effects of the interval ends are spatially well-separated. 
We demonstrate this in Figure~\ref{_cycles}, where we plot $\zeta(\mathbf{r})$ for $F$ chosen from the sets (\ref{y=0_constructive_interference}) and (\ref{y=0_destructive_interference}) (second and fourth column) 
of special Froude numbers. These are supplemented by plots for generic $F$ (first and third columns) in order to emphasize the effects.

The dd interference effects could be "unmasked" with plots of $|\zeta(\mathbf{r})|$ as in Figure~\ref{1stcycle}, but these are omitted as they do not contain any new information. In Figure~\ref{_cycles}, we overlay only the Kelvin ray $-y/(x+a/2)=1/2\sqrt{2}$ originating at the segment end as a visual aid.

The tt interference effects are more pronounced than
in Figure \ref{1stcycle}.
The waves are largest at unfavorable Froude numbers, $F=0.326, 0.252$ and $0.213$ (\ref{y=0_constructive_interference}) (second column), and smallest at favorable ones, $F=0.282, 0.230$ and $0.199$ (\ref{y=0_destructive_interference}) (fourth column).  Cyclic changes in the wake patterns
with decreasing $F$ are also visible.

It is now certain that each wake pattern is a result of superposition of the two opposite effect wakes originating at the interval edges:  the water at the front of the segment, $x=a/2$, is pulled up while at its end, $x=-a/2$, it is pushed down.  The part of the wake outside the $|y/(x+a/2)| =1/2\sqrt{2}$ wedge is solely due to the Kelvin-like wake originating at the segment's front and resembling $k_{max}=64$ wake of Figure \ref{sharp}. 

We now discuss Figure 1a of Ref.\cite{RM}, an image of a wake due to a cargo ship characterized by the Froude number $F\simeq0.15$.  It resembles the images %shown 
in Figure \ref{_cycles} with two sources, at the bow and stern, determining overall wake pattern.  The apparent wake angle that the authors associate with this Froude number corresponds to Kelvin-like wake originating at the bow of the ship.  It is then no surprise that this angle is the same as Kelvin's despite the fact that the Froude number is very small.  There also is little evidence of waves in the part of the wake where one would expect to see the two-point interference effects:  Figure 1a of Ref. \cite{RM} is very similar to the last column of images in Figure  \ref{_cycles}.  This is also not surprising as $F\simeq0.15$ is very close to the favorable Froude number $F_{7}$ given by Eq.(\ref{y=0_destructive_interference}) when the destructive tt interference effects drastically diminish the waves along the central line $y=0$.

\section{Conclusions}

Our analysis demonstrates that there is much more to 
wake patterns than apparent wake angle. For a given shape of the effective pressure disturbance, appearance of the patterns is solely determined by the Froude number $F$.
The Kelvin limit, $F=\infty$, is still relevant
as it contains physical ingredients needed to tackle the case of $F$ finite.  We find that there are at least two qualitatively different classes of wake patterns depending on whether the effective pressure disturbance is smooth or sharp (meaning it has edges or corners).  In the latter case, 
the question of how it is possible to have wakes bounded by the Kelvin angle for $F$ small is resolved by the presence of several relevant wake angles. 

In the "smooth" case, as $F$ varies from large to small values, finite size effects progressively suppress large wave vector modes. This leads to two types of wake patterns, Figure \ref{wakes}b and c, and eventually to complete elimination of the wake. Due to the non-negligible effect of large wave vector modes 
for small $F$, a wake consisting of only transverse wavefronts may be confined by the Kelvin angle if the source is strongly anisotropic (see Section VC2). Wakes with the geometry of Figure \ref{wakes}c are expected to be more elusive as they exist in a narrow range of the Froude numbers $F_{1}\lesssim F \lesssim F_{2}$ and have smaller magnitude.         

While the case of smooth pressure sources is largely understood, more work is needed to fully understand wake patterns generated by effective pressure disturbances with sharp boundaries in view of their relevance to actual ship wakes. Here we find parallel with optics to be particularly illuminating.  Indeed, we  demonstrated that the case of a uniform pressure segment with two sharp corners may be viewed as a version of Young's two-slit interference setup for gravity waves.  While Noblesse \textit{et al.} \cite{interference} started from this conjecture and then solved the problem at the level of geometrical optics, our analysis justified that approach and elevated understanding to a level comparable to that of physical optics.  The presence of two corners, which play the role of the two point sources in Young's two-slit interference setup, simplifies physical interpretation of the wake pattern. We expect that whenever the boundary has corners, they will assume the role of Kelvin-like point sources.  

However, this reasoning will not apply to pressure sources with sharp boundaries that do not have corners. An example of such a boundary is a disk of fixed radius and uniform pressure.
Understanding wake patterns due to sharp boundaries without corners is important both for applications and at the level of principle.  Indeed, the corner is a mathematical idealization that makes it possible to employ the intuition from the Kelvin point-source problem. But what allows us to approximate part of the boundary with a corner? The approximation seems plausible if the local radius of curvature of the boundary is significantly smaller than the Kelvin length (\ref{scale}). Evaluating this expectation requires separate investigation.  A related problem consists in "bridging" the cases of smooth and sharp disturbances when the pressure distribution interpolates between the smooth and sharp limits, as a function of some parameter. Progress in this direction was recently reported by Wu \textit{et al.} \cite{Wu} who studied a combined effect of the submergence depth and the hull length for a fully submerged slender body.          

\section*{Numerical Methods}

All wake integrals considered in this paper were evaluated numerically via a fast Fourier transform algorithm \cite{FFT};  a sample code is available online \cite{Jonathan_code}.   The integration region was chosen to be a $2k_{max}\times 2k_{max}$ square centered at the origin with sides parallel to the $k_{x,y}$ axes. In all of the wake images presented below, the computation parameters $k_{max}$ and the integration step $\Delta k$ were chosen to maximize the resulting image resolution for the desired real space range $x_{max}$. It can be shown for a fast Fourier transform on an array of size $N\times N$ that the resolutions in real and the wave vector space, $\Delta x$ and $\Delta k$, respectively,  obey the uncertainty relation $\Delta x \Delta k \simeq 1/N$. Here $N=k_{max}/\Delta k = x_{max}/\Delta x$, and in practice the parameter $N$ was limited by available computer memory. So at fixed $N$, one finds that $\Delta x k_{max} \simeq \Delta k x_{max} \simeq 1$. For each image, $\Delta k$ was chosen to be the smallest value compatible with the desired $x_{max}$, so that $k_{max}$ and consequentially the real space image resolution could be maximized.

For viewing purposes, the water height values, $\zeta(\textbf{r})$, were adjusted to be centered around zero. In order to prevent very large response at the origin from dominating the color scale and obscuring other wake features, a bound was applied such that all the values satisfying $|\zeta(\textbf{r})|>\zeta_{max}$ were changed to $\pm\zeta_{max}$. The exact value of $\zeta_{max}$ was determined individually for each wake image, with the goal being to have it small enough to show all of the wake features without it being so small that slight variations in the background became visible. Finally, the values were rescaled to fit on the range $[-1, 1]$.

\section{Acknowledgements}

We are grateful to I. Klich, A. P. Levanyuk and I. Shlosman for their interest in our work.

\end{document}